\documentclass[sn-nature]{sn-jnl}

\usepackage{graphicx}%
\usepackage{multirow}%
\usepackage{amsmath,amssymb,amsfonts}%
\usepackage{amsthm}%
\usepackage{mathrsfs}%
\usepackage[title]{appendix}%
\usepackage{xcolor}%
\usepackage{textcomp}%
\usepackage{manyfoot}%
\usepackage{booktabs}%
\usepackage{algorithm}%
\usepackage{algorithmicx}%
\usepackage{algpseudocode}%
\usepackage{listings}%

\theoremstyle{thmstyleone}%
\theoremstyle{thmstyletwo}%

\theoremstyle{thmstylethree}%

\begin{document}

\title[Critical assessment of water enthalpy characterization through dark environment evaporation]{Critical assessment of water enthalpy characterization through dark environment evaporation}


\author[1]{\fnm{Andrew} \sur{Caratenuto}}

\author[1,2]{\fnm{Yi} \sur{Zheng}}\email{y.zheng@northeastern.edu}

\affil[1]{\orgdiv{Department of Mechanical and Industrial Engineering}, \orgname{Northeastern University}, \orgaddress{\city{Boston}, \state{Massachusetts}, \country{USA}}}

\affil[2]{\orgdiv{Department of Chemical Engineering}, \orgname{Northeastern University}, \orgaddress{\city{Boston}, \state{Massachusetts}, \country{USA}}}


\abstract{Comparative evaporation rate testing in the absence of solar irradiation is a widely adopted method for establishing and characterizing a reduced vaporization enthalpy of water within an interfacial solar evaporator. However, the assumption of equal energy input between cases is not generally valid, and renders this characterization method misleading. Larger evaporation rates result from interfacial evaporators in dark conditions, mostly due to expanded surface areas. This causes greater evaporation rates, more evaporative cooling, and larger temperature differentials with the environment. We provide theoretical and experimental evidence to prove that the observed temperature differences are sufficient to cause large differences in environmental energy input, such that an equal energy input cannot be assumed. Evaluation of our experimental data within a transient analytical model enables quantification of these energy input differences and calculation of the vaporization enthalpy. These results show that differences in evaporation rate correspond to differences in energy input, and that vaporization enthalpy is not reduced below the theoretical value. The magnitude of increased energy input also agrees well with the increase of dark environment evaporation rates typically reported for interfacial evaporators, providing an alternative explanation to vaporization enthalpy reduction for previous literature results. Further, we exemplify that the results of this characterization method contradict differential scanning calorimetry results, which are often used to make conclusions about vaporization enthalpy reduction and water state modification within evaporator materials. Thus, we conclude that this method should not be used to make conclusions about vaporization enthalpy reductions within interfacial evaporator materials. These results emphasize that the current understanding of vaporization enthalpy reduction within evaporator materials requires re-evaluation.}

\keywords{Solar evaporation, desalination, reduced vaporization enthalpy, interfacial evaporation, dark environment}

\maketitle

\section*{Introduction}\label{sec1}
With the global water crisis becoming increasingly pervasive, research into effective mitigation methods has become very popular in recent years. Within this field, a boom of publications has emerged in the sub-field of interfacial solar evaporation, whereby solar irradiance can produce clean water with high efficiency via thermal evaporation \cite{wei_water_2023}. A great number of interfacial evaporation devices have been demonstrated for this purpose, designed to float on the surface of non-purified brine water, absorb solar irradiance, and continuously evaporate the impure water \cite{tao_solar-driven_2018}. Once evaporated, the purified water can be condensed for use, typically within some variation of a solar still.

To enhance water yield, it is highly desirable to maximize the efficiency of the evaporation process, typically by reducing parasitic energy losses to the environment. For an interfacial evaporator under solar irradiance, the steady-state evaporative heat flux can be described by

\begin{equation}
     q_{evap} = \dot{m}h_{fg} = \eta q_{sol},
\end{equation}
where $\dot{m}$, $h_{fg}$, $\eta$, and $q_{sol}$ are the evaporation rate, water phase change enthalpy, evaporation efficiency, and incident solar power, respectively. The evaporation efficiency describes the amount of incident solar irradiance used for the evaporation process, which may be reduced below 100\% by parasitic heat losses from convection, radiation, and conduction \cite{tao_solar-driven_2018}.

In an ideal case under 1 sun (1000 W m$^{-2}$) irradiance, the maximum theoretically achievable evaporation rate is approximately 1.5 kg m$^{-2}$ h$^{-1}$. Despite this, many researchers have obtained evaporation rates far above this limit, even without fully isolating the evaporation system from parasitic heat losses \cite{wei_water_2023,zhao_highly_2018,zhou_architecting_2019,caratenuto_adobe_2023,guan_sustainable_2020,liu_efficient_2022,tian_farm-waste-derived_2021,yang_flatband_2023,li_tailorable_2023}.

The typical explanation for this phenomena, introduced by Prof. Guihua Yu and colleagues in 2018, is that the vaporization enthalpy of water within the interfacial evaporators is reduced, thereby reducing the energy required to vaporize a given mass of water and allowing evaporation rates in excess of the theoretical limit \cite{zhao_highly_2018}. The proposed mechanism for this enthalpy reduction involves a weakening of the hydrogen bond network of water (or ``water activation"), resulting from interactions with the hydrophilic porous structure of the interfacial evaporator (usually a hydrogel). They further postulate that water, in this modified state, may evaporate more readily in the form of clusters, with some hydrogen bonds still intact \cite{zhao_highly_2018,zhou_architecting_2019}. After this point, the clusters may break apart outside of the system (i.e., above the evaporator), reducing the energy requirement for the evaporator itself.

A variety of characterization methods were employed to defend this hypothesis, many of which are frequently used in contemporary studies to justify evaporation rates above the theoretical limit and prove that the vaporization enthalpy of water has been reduced. One of the most commonly-used methods involves evaluating evaporation rates in the absence of solar irradiance (i.e., in a dark environment). In the dark environment evaporation test, the mass loss of identical water-filled beakers -- one of pure water, and one with the interfacial evaporator placed on top -- is monitored over an extended duration. With the assumption that the energy input from the environment (which facilitates pseudo-steady evaporation over the test period) is equal between the two beakers during the process, one can conclude that 

\begin{equation}
     (\dot{m}h_{fg})_{H_2O} = (\dot{m}h_{fg})_{evap}.
     \label{eqMass}
\end{equation}
Thus, if the evaporation rate of the interfacial evaporator is higher than that of pure water in these conditions, this method offers a simple indicator that the vaporization enthalpy of water within the evaporator must be reduced \cite{zhao_highly_2018,wei_water_2023}.

\begin{figure} 
    \centering
    \includegraphics[width=0.8\textwidth]{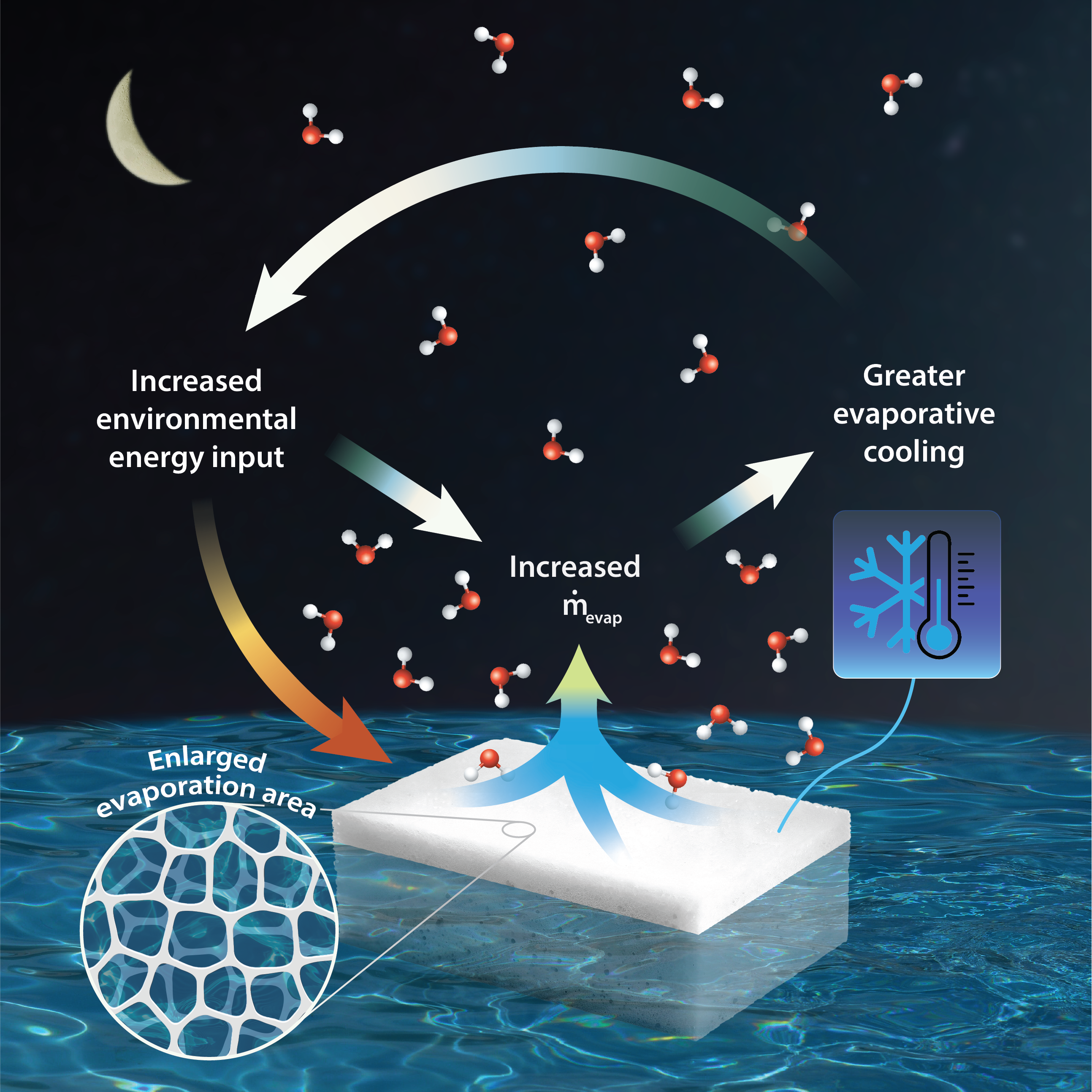}
    \caption{\textbf{Mechanism of enhanced dark environment evaporation from an interfacial evaporator with enlarged surface area.} The larger surface area increases evaporation, inducing greater evaporative cooling, thus lowering the temperature of the evaporator and allowing it to harvest additional energy from the environment.}
    \label{f1}
\end{figure}

However, deeper investigation reveals that this method produces misleading results that should not be used to justify a reduced vaporization enthalpy, due to the invalidity of the equal energy input assumption. In this work, we show both theoretically and experimentally that differences in the dark environment evaporation rate are primarily due to differences in evaporation surface area, which invalidates the assumption of equal energy input (Fig. \ref{f1}). It is illustrated that energy input discrepancies due to differences in system temperatures are proportional to differences in evaporation rates. Modeling results show good agreement with the experiment, supporting these findings. Finally, we provide data to show that differential scanning calorimetry methods, which are commonly used to characterize vaporization enthalpy, are not in agreement with the results of the dark environment test.

\section*{Theoretical Analysis and Hypothesis}\label{sec2}
Left at rest in ambient conditions, water in a beaker will spontaneously evaporate as liquid molecules at the interface gain enough energy to overcome surface binding energy. In doing so, energy exits the water system through the latent heat of the exiting water vapor. This energy output manifests as a sensible heat reduction in the remaining liquid water, cooling the water. As the water temperature falls below the ambient temperature via evaporative cooling, the water will begin to receive energy from the environment due to the temperature differential, compensating for evaporative energy losses. Pseudo-steady evaporation is achieved when the net energy input from the environment is equal to the energy output due to evaporation \cite{incropera_fundamentals_2007,li_measuring_2019}.

It is well known that evaporation rates are heavily dependent on the interfacial area, and that enlarging the evaporation surface will accelerate mass transfer. Many interfacial evaporators possess significantly enlarged surface areas, typically obtained via an internal porous structure, which intrinsically enhances the natural evaporation rate \cite{liu_easy--fabricate_2021,bongarala_figure_2022}. Furthermore, other factors such as pressure modification within a porous structure or the modification of surface energies may further enhance evaporation \cite{wei_water_2023,yu_enhanced_2022}. All of these factors are generally desirable for increasing the evaporation rate, which can be clearly shown by an enhancement of evaporation under dark conditions.

However, these evaporation-enhancing factors have not been shown to inherently reduce the total energy requirement of evaporation. Under steady evaporation conditions, evaporative cooling must be balanced by energy from the environment via radiation, convection, and conduction. Energy input from any of these mechanisms is proportional to the temperature difference between the ambient environment and the surface(s) of the evaporation system. Therefore, for a fixed ambient temperature, an evaporation system with lower temperatures at steady state will receive a greater energy input from the environment.

In the context of a typical comparative dark environment test, the enhanced natural evaporation of the interfacial evaporator will induce greater cooling, which in turn induces a greater environmental energy input. This temperature difference may appear small -- often less than 1 °C between the water and evaporator systems -- yet as will be shown, can cause appreciable discrepancies between the energy input of the two systems. As a result, the interfacial evaporation device receives significantly more energy input, and can sustain enhanced evaporation rates in comparison to water without relying on the hypothesis of reduced vaporization enthalpy. In short, enhanced evaporation resulting from the properties of the interfacial evaporator produces a greater evaporative cooling effect, which enables greater environmental heat fluxes that can maintain larger evaporation rates (Fig. \ref{f1}).

\subsection*{Theoretical heat transfer model}\label{sec3}
To prove that the temperature differences which result from disparate evaporation rates are not negligible in a comparative dark environment evaporation test, theoretical calculations are performed. Consider a beaker of diameter $d$ and height $H$ filled with water, placed on a thick piece of insulating foam. In general, convective and radiative transfer between a surface of the beaker and the environment can be described by the following equations \cite{incropera_fundamentals_2007}:

\begin{equation}
    q_{conv} = A_ih_i(T_{\infty}-T_i)
        \label{convection_HT}
\end{equation}

\begin{equation}
    q_{rad} = A_i\varepsilon_i\sigma(T_{\infty}^4 - T_i^4).
        \label{rad_HT}
\end{equation}

In these equations, the subscript $i$ is used to designate the appropriate surface of the beaker. $A$ is the area of the surface, $h$ is the convection coefficient, $T_{\infty}$ is the ambient temperature, $T$ is the beaker surface temperature, $\varepsilon$ is the emissivity of the surface, and $\sigma$ is the Stefan-Boltzmann constant. The Nusselt number is used to calculate natural convection coefficient for the horizontal top surface and vertical walls of the beaker, respectively, as \cite{incropera_fundamentals_2007}:

\begin{equation}
  \begin{aligned}
    & Nu_L = 0.54Ra_L^{1/4} \:\: for \:\: Ra_L < 10^7\\
      & Nu_L = 0.15Ra_L^{1/3} \:\: for \:\: Ra_L \geq 10^7\\
              \label{nu_1}
  \end{aligned}
\end{equation}

and

\begin{equation}
    Nu_L = \bigg\{{0.825+\frac{0.387Ra_L^{\frac{1}{6}}}{[1+(\frac{0.492}{Pr})^{\frac{9}{16}}]^{\frac{8}{27}}}}\bigg\}^2.
\end{equation}

Air properties are evaluated at the average temperature between the ambient and the surface. The wall of the beaker is assumed to have negligible thermal resistance due to its low thickness. The beaker is also assumed to be insulated on the side and/or bottom surfaces (depending on the case). For simplicity, steady state is assumed, such that changes in sensible heat can be neglected. The accuracy of these assumptions will be revisited in later sections with respect to the experimental results.

The emissivities of the water, beaker, and evaporator are taken as 0.96, 0.95, and 0.93, respectively. The diameter and height of the beaker are 41 and 40 mm, respectively, and the Stefan-Boltzmann constant is taken as 5.67$\times$10$^{-8}$ W m$^{-2}$ $K^{-4}$. When the side walls of the beaker are considered, the side wall temperature is assumed to be 0.3 °C above the evaporation surface temperature for the water and 1.1 °C for the evaporator, based on our experimental observations and previous results in the literature. The calculation also limits the wall temperatures from exceeding the ambient temperature and limits the evaporator wall from exceeding the water wall temperature, as either case is not likely to occur in a practical scenario.

Using these methods, the environmental energy input for a beaker undergoing steady evaporation is quantified. By modeling and comparing beakers with different surface temperatures in the same ambient conditions, the energy impact of these temperature differentials is quantified. In this way, an input energy enhancement factor, which describes the excess energy input brought on by lower system temperatures (as may result due to the enhanced evaporation of an interfacial evaporator), can be defined as

\begin{equation}
    f_{in,evap} = \frac{U_{in,evap}}{U_{in,H_2O}},
\end{equation}
where $U_{in,evap}$ and $U_{in,H_2O}$ represent the net input energy from the environment to the interfacial evaporator and water system, respectively. Thus, the energy enhancement factor provides a reasonable estimate for the energy input differences between the water and evaporator systems during a comparative dark environment evaporation test.

\begin{figure} 
    \centering
    \includegraphics[width=\textwidth]{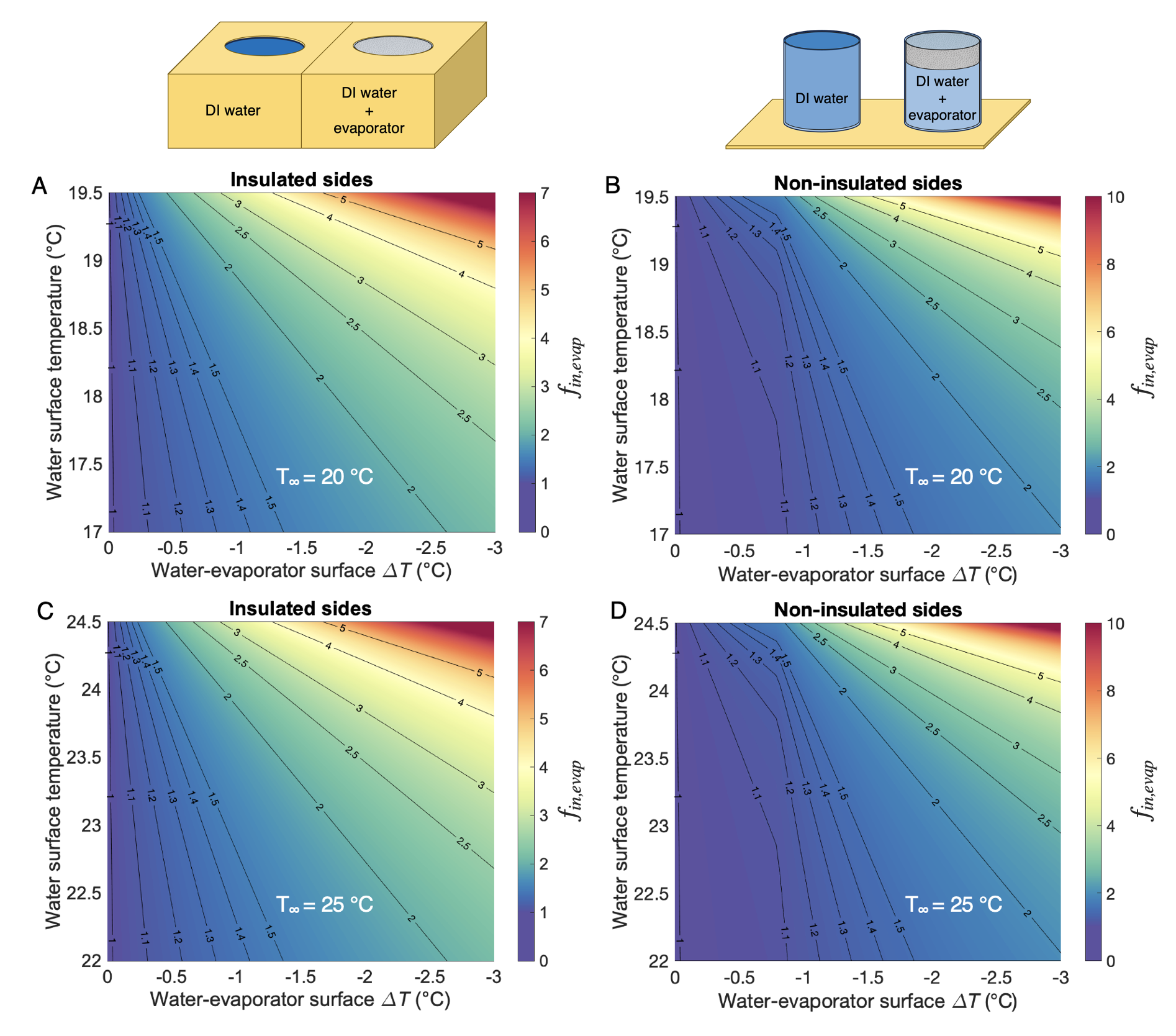}
    \caption{\textbf{Theoretical modeling results.} Input energy enhancement factor $f_{in,evap}$ as a function of water and interfacial evaporator surface temperatures under a fixed ambient temperature $T_{\infty}$. (A) and (B) present results for cases of insulated and non-insulated side walls of the beakers, respectively, at $T_{\infty}$ = 20 °C. (C) and (D) show the same two cases with an ambient temperature of $T_{\infty}$ = 25 °C}
    \label{f2}
\end{figure}

\subsection*{Analytical results}\label{sec4}
The results of these theoretical analyses are shown in Fig. \ref{f2}. First, an ambient temperature of 20 °C is studied, with water surface temperatures ranging from 17 -- 19.5 °C (y-axis). The interfacial evaporator beaker is designated with a temperature difference from 0 to 3 °C below that of the pure water (x-axis). From Fig. \ref{f1}a, it is immediately apparent that even small surface temperature differences can cause quite large differences in energy input. The energy input to the evaporator system can exceed 1.5 times that of the water system even with a temperature differential less than 1 °C. At greater differentials, the evaporator system can easily obtain 2-3 times as much energy from the environment as the pure water system.

As expected, the energy input enhancement factor $f_{in,evap}$ is heavily dependent on the chosen temperature differential between the water and evaporator systems, as all heat transfer components are directly proportional to this quantity. In addition, if the chosen water temperature is closer to the ambient temperature, the relative energy gain from the corresponding evaporator case is larger. This is because the ratio of the evaporator-ambient temperature difference and the water-ambient temperature difference becomes quite large in these cases.

When the sides of the beaker are not insulated (i.e., convective and radiative heat transfer from the sides is considered), the evaporator generally receives less heat transfer for a given surface temperature differential, as seen in Fig. \ref{f2}b. This is due to the fact that the side wall temperatures are higher than those of the evaporation surface, and that the evaporator wall temperature is set with a larger differential than that of the water. As a result, the difference in energy input for a given surface temperature differential is slightly subdued. In addition, the choice of ambient temperature has a minimal impact, which is expected (Fig. \ref{f2}c--d). The ambient temperature essentially only modifies the convection coefficients through the air properties, and in the range of temperatures modeled here, air does not experience property changes that majorly impact convection. Thus, the impact of ambient temperature is minimal in comparison to the temperature differentials, which emerge as the obvious driving factors. Together, these observations indicate that the conclusions are largely setup and environment-independent, and thus transferable to comparative dark environment tests across different laboratory environments and for various common beaker setup choices (insulated or non-insulated).

The most important takeaway of the analytical results concerns the magnitude of the differences in energy input. The model indicates that temperature differences in the range of 0 -- 3 °C may provide the evaporator with far more energy than the water system, more than doubling its energy input in many cases. Thus, this enhanced energy input would explain evaporation rates of interfacial evaporators which exceed those of water by similar factors without necessitating claims of reduced vaporization enthalpy.

\begin{figure} 
    \centering
    \includegraphics[width=\textwidth]{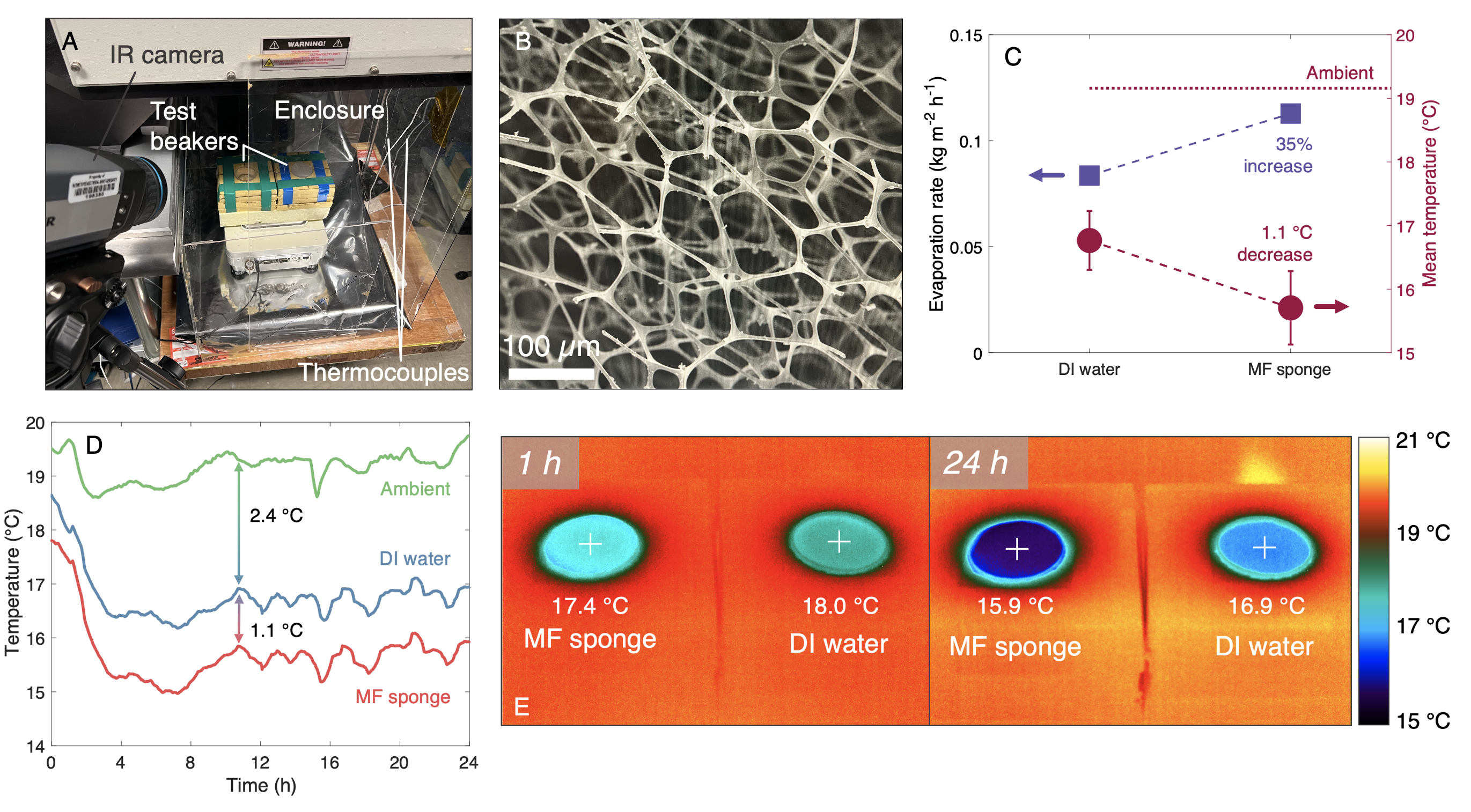}
    \caption{\textbf{Experimental results for comparative dark environment evaporation test.} (A) Dark environment system setup. (B) SEM image of the open porous structure of the MF sponge evaporator. (C) 24-h dark environment evaporation rate and average evaporation surface temperature for the DI water and MF sponge evaporation systems. (D) System temperatures during the evaporation test. (E) IR camera images of both systems during the test, showing their disparate evaporation surface temperatures.}
    \label{f3}
\end{figure}

\section*{Experimental validation}\label{sec5}
\subsection*{Evaporation tests}
To further support these claims, the comparative dark environment test is performed (Fig. \ref{f3}). The interfacial evaporator chosen for these tests is a simple melamine foam (MF) sponge, exemplified previously as an interfacial solar evaporator \cite{liu_efficient_2022}. This evaporator demonstrates superb water transport, high hydrophilicity, and has an open-porous structure with an average pore diameter of approximately 30 µm, as seen from the SEM images (Fig. \ref{f3}b and S1). Two beakers containing DI water are placed side by side in blocks of insulation, as shown in Fig. \ref{f3}a and S2. The MF sponge is placed on the water surface of the evaporator beaker, in direct contact with the DI water, with its top surface in line with the mouth of the beaker. Temperatures of each evaporation system and the environment are monitored using an IR camera and confirmed with thermocouples, and the mass of each beaker is weighed before and after 24 h to determine the mass loss by evaporation. Further details are provided in the \textit{Experimental methods} section.

The experimental results confirm our key hypothesis: the interfacial evaporator system maintains notably lower temperatures than the DI water system throughout the entirety of the test. While both systems display a fairly consistent temperature differential with the ambient after an initial transient period, the MF sponge maintains a temperature nearly 1.1 °C below that of the DI water surface under identical ambient conditions, as shown in Fig. \ref{f3}c--d. 

As expected, the use of the interfacial evaporator yields an enhanced evaporation rate, as shown in Fig. \ref{f3}c. The DI water beaker achieves an evaporation rate of 0.084 kg m$^{-2}$ h$^{-1}$ during the 24-h test. In contrast, the MF sponge achieves a 35\% higher evaporation rate over the same period (0.113 kg m$^{-2}$ h$^{-1}$). The porous structure of the MF sponge enhances the evaporation rate primarily due to its enlarged evaporation area; its pore sizes are in the double-digit micron range, which is not expected to cause any appreciable difference in internal pressure \cite{liu_efficient_2022,hou_self-assembly_2019}. However, this characteristic has no effect on the energy required for water to evaporate from its porous structure. Instead, its ability to maintain higher evaporation rates is due to its lower interfacial temperatures (clearly seen in the IR camera images of Fig. \ref{f3}e),  which allow for a greater energy input from the environment.

\subsection*{Transient model}
To support these conclusions, transient analyses are performed on the experimental temperature data. The energy balance for the entire beaker during the test can be expressed as

\begin{equation}
    q_{evap}=\dot{m}h_{fg}=q_{conv}+q_{rad}+q_{cond}+q_{sens},
    \label{ebalance}
\end{equation}
where $q_{cond}$ and $q_{sens}$ represent conductive and sensible heat contributions. This equation is consistent with earlier theoretical descriptions of the evaporation process, where energy loss from evaporation is balanced by cooling of the water itself and environmental energy inputs. This relationship enables quantification of the energy transfer components and estimation of the vaporization enthalpy of each beaker, which will further illustrate that the assumption of equal energy input is invalid, and that there is no reduction in vaporization enthalpy.

A transient, axisymmetric heat transfer model is run for each beaker using the Matlab PDE toolbox \cite{noauthor_partial_nodate}. The boundary condition at the top surface for each time step is set as the experimentally measured surface temperature of the evaporation surface. The side and bottom walls are designated with conductive heat flux boundary conditions, based on measured values of the insulation thickness and thermal conductivity, and water properties are approximated as constant due to the low range of temperatures. Further information regarding model input parameters is provided in Supplementary Note 1.

Based on these inputs, the temperature distribution inside the beaker is calculated at each time step over the 24 h test period. The solution is nonlinear, as conductive and sensible heat inputs are determined based on the resulting temperature field. Transient convective and radiative inputs at the evaporation surface are determined using Eqs. \ref{convection_HT} -- \ref{nu_1}. The ambient air temperature is used for the convection calculation, while the enclosure wall temperature is used for the radiative calculation. Thus, all contributing energy input components are determined for the experimental test.

\begin{figure} 
    \centering
    \includegraphics[width=\textwidth]{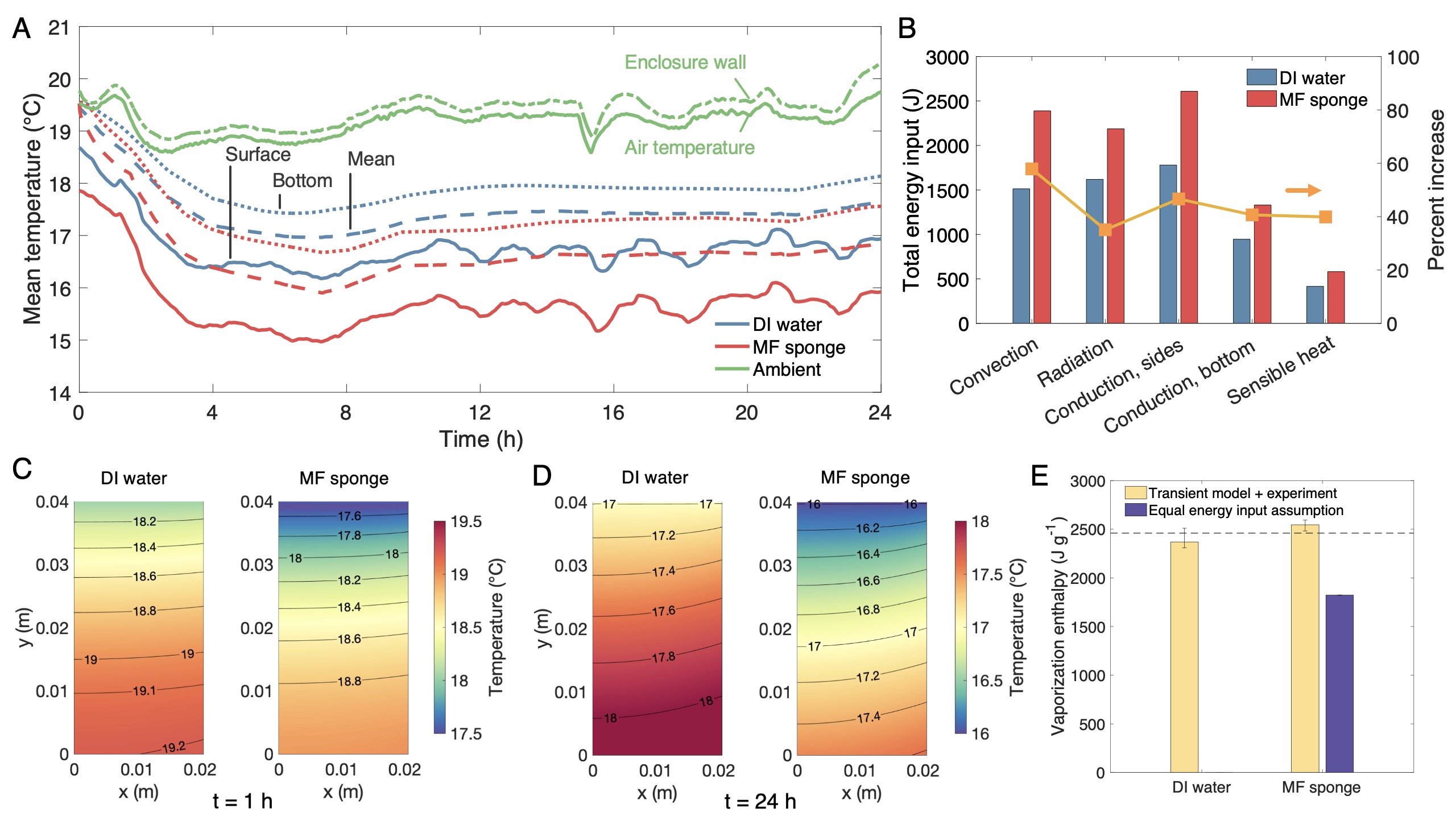}
    \caption{\textbf{Experimental analysis using transient model.} (A) Temporal data for experimental system and resulting system temperatures based on transient model. Solid, dashed and dotted lines for beaker systems represent the surface (experimental), bulk mean (model), and bottom (model) temperatures, respectively. (B) Cumulative energy input from different environmental heat transfer mechanisms over the 24-h test period. Yellow markers denote percent increase in each component when the MF sponge is used, corresponding to the right axis. Modeled axisymmetric temperature profiles within the beakers at (C) 1 h and (D) 24 h. (E) Calculated vaporization enthalpy determined from the transient model analysis. The dotted line indicates the theoretical enthalpy value of 2460 J g$^{-1}$. Error bars indicate model results based on an IR camera measurement uncertainty of $\pm$ 0.05 °C. The secondary purple bar shows the erroneous enthalpy value for the MF sponge that would be obtained if the equal energy input assumption were used.}
    \label{f4}
\end{figure}

Figure \ref{f4} provides the results of these analyses. From Fig. \ref{f4}a and c--d, the bottom and mean surface temperatures of each beaker are higher than the evaporation surface, as expected. The maximum temperature is maintained at the bottom of each beaker, as it is farthest from the cold evaporation surface while still receiving energy input from the surroundings. Model mean and bottom temperatures respond accordingly to changes in ambient and surface temperatures, yet are temporally delayed due to the large specific heat of water. At all times, the entire MF sponge beaker maintains bulk and interface temperatures below that of the DI water beaker. 

The cumulative energy input from each heat transfer mechanism is shown in Fig. \ref{f4}b. For all mechanisms, the MF sponge beaker receives significantly more energy than the DI water beaker. The lower temperatures of the MF sponge system induce greater heat transfer through an increased temperature differential with the environment, resulting in approximately 40\% more energy input for most heat transfer mechanisms. For convection, this percent increase is even larger, nearing 60\%. Convection input to the MF sponge system is enhanced not only by an increased environmental temperature differential, but also by a slight increase in the convection coefficient (Fig. S4). Even the sensible heat contribution, though small, is enhanced by the lower system temperatures of the MF beaker. Conduction from the side walls emerges as a key contributor to the energy input of both systems. However, it is important to note that the side wall area is nearly 4 times as large as the top or bottom surfaces, over which the other environmental inputs act upon. Thus, on the basis of heat flux, convection and radiation on the top evaporation surface are the largest contributors. This outcome is expected, as the largest environmental temperature differential exists at this location.

By combining these results with Eq. \ref{ebalance}, the vaporization enthalpy of each beaker can be calculated, shown in Fig. \ref{f4}e. Analysis of the water beaker yields an enthalpy value of 2370 J g$^{-1}$, while that of the MF sponge beaker is 2546 J g$^{-1}$. The values of both beakers agree with the theoretical value of 2460 J g$^{-1}$ \cite{moran_fundamentals_nodate} within less than 4\%. This result illustrates clearly that the vaporization enthalpy of water within the MF sponge is not reduced, but that the assumption of equal energy input used in the comparative dark environment test is invalid. The MF sponge receives a greater energy input based on its lower temperatures, which corresponds with great accuracy to its enhanced evaporation rate. If the equal energy input assumption were employed for this test, a vaporization enthalpy value of 1823 J g$^{-1}$ would have been established. Thus, it is easy to see how misleading results of vaporization enthalpy reduction are obtained using this method. These experiments and analyses show clearly that, when energy inputs from the environment are properly accounted for, there is no indication of vaporization enthalpy reduction. Further support for the reliability of these analyses is provided in Supplementary Note 2 and Fig. S5.

\begin{figure} 
    \centering
    \includegraphics[width=0.65\textwidth]{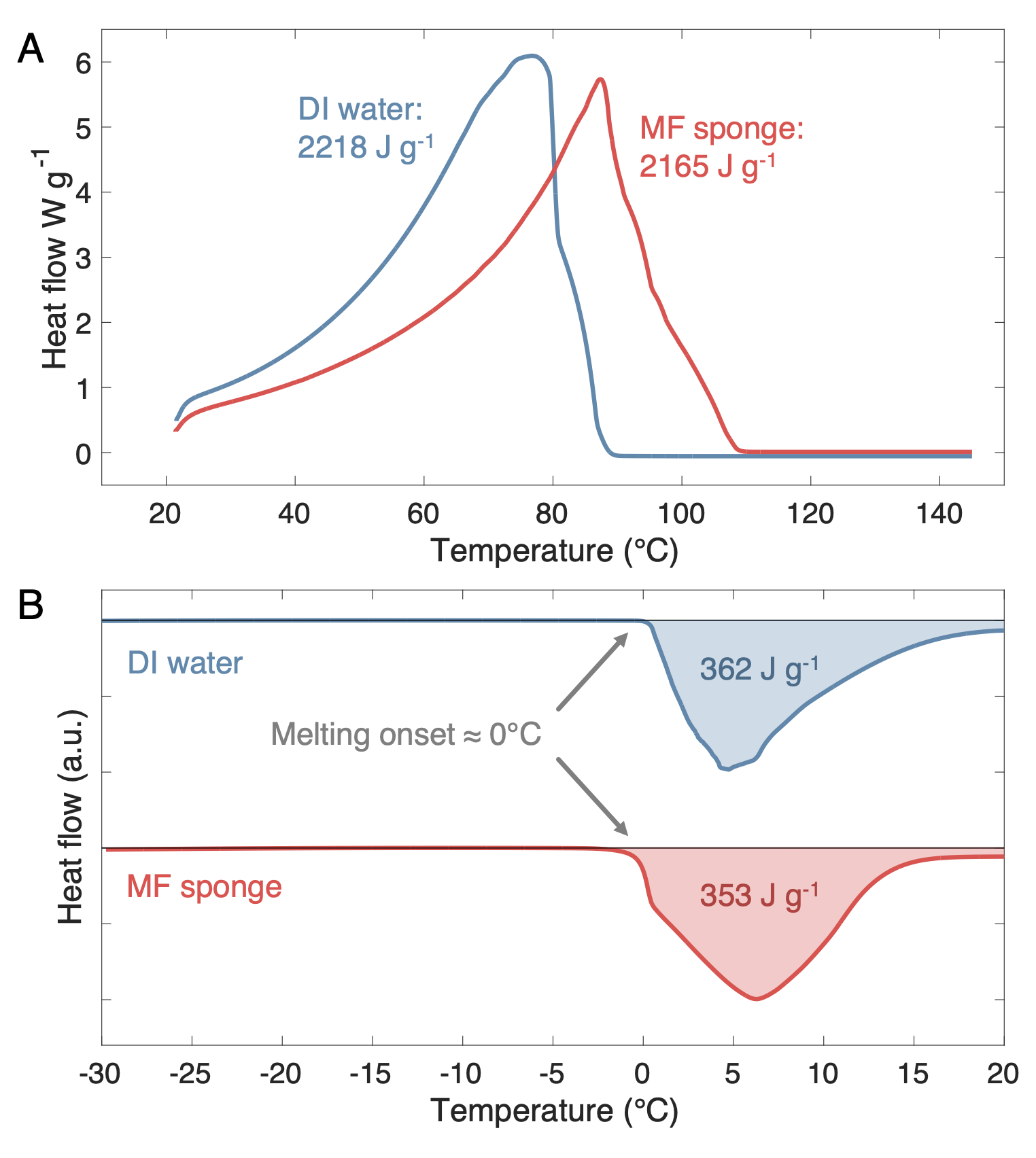}
    \caption{\textbf{Differential scanning calorimetry.} Heat flow curves for DI water and saturated MF sponge at (A) high temperatures and (B) low temperatures. Transition enthalpies, obtained through integration, are provided in the plots.}
    \label{f5}
\end{figure}

\subsection*{Differential scanning calorimetry (DSC) verification}
Finally, the MF evaporator is characterized using DSC to further support these conclusions. DSC data is used to characterize transition enthalpies, and features of the heat flow curves can be analyzed for further information \cite{wei_water_2023,zhou_architecting_2019}. DSC scans at high and low temperatures are frequently used for interfacial evaporators to prove their modified water state and/or reduced vaporization enthalpy, either on their own, or in conjunction with dark environment test data.

In Fig. \ref{f5}, the DSC scan of pure water over the boiling point yields a vaporization enthalpy value close to the theoretical value, and shows a steep characteristic decline after all water is vaporized. As can be seen in Fig. \ref{f5}a, high-temperature DSC scans for pure water and the MF sponge show very similar results. Both profiles yield vaporization enthalpy values close to the theoretical value of water. The vaporization enthalpy of water in the MF sponge decreases by only 2\% in comparison to pure water, and the two profiles have a similar characteristic steepness after the peak is reached. In contrast, interfacial evaporators with modifications in water state will show DSC enthalpy values well below that of pure water, typically showing reductions of 20\% or more (vaporization enthalpy values below 2000 J g$^{-1}$). A broader, less steep profile is usually seen as well, and heat flow activity may persist past 100 °C, both of which are attributed to the modified state of water within the evaporators \cite{zhao_highly_2018,zhou_architecting_2019,wei_water_2023}. None of these factors are visible for the MF sponge; hence, high-temperature DSC scans do not indicate any reduction in the vaporization enthalpy of water within the MF sponge. We note that the shift of the MF sponge heat flow to higher temperatures is due to a slightly larger mass of water used in this test. 

The DSC can also be used at low temperatures to gain information about the bonding state of water. Typically, for a modified water state as is commonly seen in hydrogels, the melting onset will shift well below 0 °C, and the melting enthalpy often decreases \cite{zhou_architecting_2019,wei_water_2023,nakaoki_bound_2008}. However, both the DI water and MF sponge show a melting onset very close to 0 °C (Fig. \ref{f5}b). Both melting enthalpies are close to the theoretical value (334 J g$^{-1}$), and agree with one another within less than 3\%. 

Together, these low and high temperature DSC results clearly show that neither the vaporization enthalpy nor the hydrogen bonding state of water within the MF sponge are modified with respect to those of bulk water. Despite this, the dark environment test on its own would indicate a reduced vaporization enthalpy if the assumption of equal energy input were used. We note that discrepancies between reduced enthalpy values characterized via dark environment or DSC methods are common, and are usually attributed to differences in the degree of dehydration experienced by the evaporator within each test \cite{zhao_highly_2018} . However, in such scenarios in the literature, both the dark environment and DSC tests indicate a reduced enthalpy, typically with slightly different values. In contrast, our results show no indication of a reduced enthalpy from the DSC. Thus, we posit that DSC and dark environment characterization methods must not characterize the same phenomena regarding vaporization enthalpy, as they produce contradictory results. 

\section*{Discussion}\label{sec6}
The data and analyses reported herein clearly illustrate that comparative dark environment evaporation tests are not suitable for characterizing the reduced vaporization enthalpy of an interfacial evaporator. Enlarged evaporation surface areas or other evaporator characteristics may increase the evaporation rate in a dark environment, but this does not inherently modify the energy requirement of evaporation. Rather, increased evaporative cooling allows the evaporator system to reach a lower steady state temperature, enabling it to harvest more environmental energy due to a greater temperature differential. Crucially, theoretical and experimental results show that the magnitude of this enhanced energy input is proportional to the enhanced evaporation observed in a dark environment. Furthermore, DSC results, which are commonly used to corroborate claims of reduced vaporization enthalpy in interfacial evaporators, are inconsistent with the dark environment results presented. Thus, it is concluded that comparative dark environment tests are not a valid method for establishing or quantifying a reduced vaporization enthalpy in an interfacial evaporator system, as the key assumption of equal energy input is not accurate.

To thoroughly and accurately demonstrate this concept, well-controlled experiments and detailed analytical models have been employed. However, the utility of simple theoretical calculations in demonstrating this phenomenon is also apparent. If Fig. \ref{f2} were used to analyze the experimental data given in Fig. \ref{f3}, an input energy increase of about 45\% would have been noted, corresponding to a MF sponge evaporation rate about 45\% larger than that of DI water. This is an overestimate, largely due to sensible heat and conductive contributions which are not accounted for. However, as most energy input mechanisms increase by roughly the same magnitude when the MF sponge is used, this simple method does offer a relatively close approximation, useful for benchmarking other reports in the literature. 

Previous accounts of reduced enthalpy from dark environment evaporation tests are widespread in the literature. Table \ref{tbl:1} shows literature reports of reduced enthalpy obtained via the dark environment evaporation method ($h_{fg,evap}$). These values range from approximately 1000 -- 2000 J g$^{-1}$. Based on Eq. \ref{eqMass}, the enhancement factor of evaporation from the interfacial evaporator can be defined as $f_{\dot{m}_{evap}}=h_{fg,evap}/h_{fg,H_2O}$. The reported values of $h_{fg,evap}$ correspond to $f_{\dot{m}_{evap}}$ values in the 1 -- 2.5 range.

\begin{table*}
\small
  \caption{Literature reports of reduced vaporization enthalpy (and associated ratio of evaporation increase) characterized using the dark environment evaporation method.}
  \label{tbl:1}
  \begin{tabular*}{\textwidth}{@{\extracolsep{\fill}}llll}
    \hline
    Material & $h_{fg,evap} (J\ g^{-1})$ & $f_{\dot{m}_{evap}}$ \\
    \hline
    Carbon cloth  \cite{yu_enhanced_2022} & 2069 & 1.19 \\
    PPy/Carbon cloth  \cite{yu_enhanced_2022} & 1715 & 1.43 \\
    Titanium suboxide powder  \cite{yang_flatband_2023} & - & 2.33 \\
    PVA-PPy hydrogel  \cite{zhao_highly_2018} & 1377 & 1.75 \\
    PVA-chitosan hydrogel  \cite{zhou_architecting_2019} & 1030 & 2.38 \\
    Adobe brick  \cite{caratenuto_adobe_2023} & 1153 & 2.06 \\
    Carbonized manure  \cite{tian_farm-waste-derived_2021} & 1276 & 2.02 \\
    MF sponge + Black 3.0  \cite{liu_efficient_2022} & 1377 & 1.75 \\
    Carbonized wood  \cite{guan_sustainable_2020} & 2069 & 1.19 \\
    Carbon foam  \cite{serrano_utilization_2022} & 2013 & 1.22 \\
    Lignocellulose aerogel  \cite{li_tailorable_2023} & 1434 & 1.71 \\

    \hline
  \end{tabular*}
  
\end{table*}

Looking back at the model results of Fig. \ref{f2}, these accounts of reduced enthalpy could easily be explained by small variations in interfacial temperature during the dark environment tests as opposed to differences in vaporization enthalpy. It is difficult to assess past reports in the context of our model, as experimental temperatures during dark environment tests are scarcely reported. However, recent works show that water-evaporator surface temperature differences up to nearly 3 °C are certainly reasonable \cite{yang_flatband_2023}. In addition, many of these comparative tests are not performed concurrently, meaning that the water and evaporator systems may experience differences in ambient temperature and humidity. This can lead to even further discrepancies in the environmental energy input, especially due to the influence of humidity. Considering these factors, the vaporization enthalpy reductions that are typically reported using the dark environment evaporation rate test correspond well with the magnitude of model-predicted excess energy input for a reasonable range of temperature differentials. Although some references report higher ratios of evaporation increase than our experimental results, we attribute this to the fact that our tests, being performed concurrently, maintain better control of ambient conditions than many literature reports, which may perform evaporation tests at separate times. The MF sponge also has a limited surface area enhancement due to its relatively large pore sizes, and is unlikely to have any impact on internal pressure \cite{liu_efficient_2022,bongarala_figure_2022}. Considering these factors, differences in environmental energy input provide a far more likely explanation for the disparate evaporation rates observed in comparative dark environment tests throughout the literature, as opposed to a reduction in vaporization enthalpy. 

With the context of these experimental results, it is important to comment on the controversial role of the dark environment evaporation rate in the analysis of a solar-driven evaporation system \cite{luo_energy_2021,li_measuring_2019}. Often, the dark environment evaporation rate is subtracted from the evaporation rate under solar irradiance to obtain a dark-excluded evaporation rate. The goal of this is to isolate the impact of incident irradiation on the evaporation system from that of natural evaporation. However, we posit that this is unnecessary, especially when calculating the evaporation efficiency. Pseudo-steady dark environment evaporation is maintained due to environmental energy input, which stems from a negative temperature differential between the evaporation system and the environment. This environmental energy input typically does not manifest during an illuminated solar evaporation test, as this negative temperature differential does not exist. If all regions of the evaporator itself are elevated above the ambient temperature, there will not be energy input from the surroundings. Thus, in a properly isolated evaporation rate test, the evaporator should not receive excess energy from the environment, and there is no need to subtract the evaporation rate from such a source, as noted by Li et al \cite{li_measuring_2019}. Exceptions to this include cases with unique geometries designed to harvest environmental energy input \cite{tang_realization_2020,liu_easy--fabricate_2021}. However, even for these exceptions, the surrounding energy input is highly dependent on ambient conditions (temperature and humidity), which influence input heat transfer and evaporation rate. In an illuminated evaporation rate test, these conditions will certainly be modified, and the input energy cannot be assumed to be the same as that of a dark environment test. While the dark environment rate may be subtracted in these cases for an estimate of the pure, solar-driven evaporation, this method provides, at best, only a conservative estimate \cite{li_measuring_2019}. For these reasons, simply subtracting the dark environment evaporation rate from the solar-driven evaporation rate is inaccurate and unnecessary for the calculation of the solar-to-vapor conversion efficiency in most systems, as the environmental energy input which sustains dark evaporation is not identical (and in most cases does not manifest) in a typical illuminated test.

Finally, recent discussions in the literature call into question our current understanding of reduced enthalpy of evaporation during continuous interfacial solar evaporation processes. A modification of the water state clearly manifests in many materials (most notably hydrogels), and produces a reduced value of the DSC-measured enthalpy along with solar-driven evaporation rates above the theoretical limit \cite{wei_water_2023}. However, when considering the energy balance of the entire system, the enthalpy of the water prior to inundation within the hydrogel is unchanged from its theoretical bulk value. Thus, in a continuous system, energy should be consumed to raise the water's enthalpy from the bulk value to the modified value when it is absorbed into the hydrogel - which must consume energy, as noted by Ducker \cite{ducker_decreasing_2023}. For this reason, it seems impossible that reductions in water vaporization enthalpy would manifest in a continuous evaporation system if the phenomena were based purely on a modification of the bonding state of hydrogel-absorbed water. Our experimental data lends support to this theory. Firstly, it shows that typical claims of enthalpy reduction in a continuous evaporation test are likely invalid, offering the alternative explanation of energy input discrepancies. In addition, it illustrates the contradictory nature of continuous (dark environment) and non-continuous (DSC) enthalpy characterization methods. Rather, alternative theories which involve extra energy input coming from outside of the evaporation system would provide a more reasonable explanation, such as those which posit the breakup of released water clusters outside of the evaporator system \cite{tu_plausible_2023}. This would also help explain how many evaporation materials besides hydrogels are capable of breaking the theoretical limit even without a modification of the water state. Further efforts by the scientific community are vital to advance our understanding of the complex vaporization phenomena which occur during solar-driven evaporation. A deeper understanding of how the vaporization energy requirement may be effectively reduced, materials which facilitate these phenomena, and suitable characterization methods will provide significant progress towards global sustainability.

\section*{Conclusion}\label{sec7}
In this work, a critical assessment of reduced enthalpy characterization through dark environment evaporation testing is performed. The assumption of equal energy input between a water system and an interfacial evaporator system is analyzed both theoretically and experimentally. In both cases, the environmental energy input to the interfacial evaporation system is shown to significantly exceed that of the water system. The excess energy input within the experimental test is calculated using a transient analytical model, and it is found that the enhancement in evaporation for the interfacial evaporator agrees with the enhancement of energy input due to lower evaporator temperatures. This data illustrates that the assumption of equal energy input is not valid, and that reduced values of enthalpy obtained using this characterization method are not accurate. Recent uses of the dark environment characterization method in the literature are analyzed with respect to these results, and it is found that the magnitude of previously reported dark environment data matches well with our explanation of excess energy input. Discrepancies between differential scanning calorimetry methods are also discussed, emphasizing that the current understanding of vaporization enthalpy reduction in interfacial evaporation systems requires clarification and deeper investigation.

\subsubsection*{Experimental methods}\label{expmeth}
Commercially-available melamine foam sponges are obtained from South Street Designs Company (UPC: 089902974060). The sponge is rinsed 3 times with DI water and ethanol before use. It is cut to a 41 mm diameter to fit within the test beaker during evaporation tests. Evaporation tests were performed over a test period of 24 h in a closed darkroom environment. Two polypropylene beakers containing DI water were placed side-by-side, undisturbed during the entire test period. The MF sponge is placed on the water surface of the evaporator beaker, in direct contact with the DI water. It is pushed into the beaker until its top surface is in line with the mouth of the beaker. The beakers were surrounded on the sides by 42 mm of polyvinyl chloride (PVC) insulation foam (measured thermal conductivity of 0.065 W m$^{-1}$ K$^{-1}$). The beakers were placed on top of 38 mm of polystyrene (PS) insulation foam (measured thermal conductivity of 0.047 W m$^{-1}$ K$^{-1}$) and another 12.7 mm of PVC insulation foam. Insulation boards were firmly secured to one another to diminish contact resistance. Temperatures were monitored using the FLIR A655C thermal camera with a resolution of 640 × 480 using a 25° lens. Temperatures for evaporator system surfaces were taken as the average over each respective surface. Transparent acrylic sheets were positioned around the setup to reduce the impact of ambient temperature and wind fluctuations during the test. This enclosure is not air-tight, so that evaporating water vapor will not accumulate locally. Ambient temperature was monitored using a thermocouple positioned inside of the wind-covered area. An additional thermocouple was secured to the inside of one acrylic sheet to monitor the temperature of the walls inside the enclosure. The enclosure wall temperature was used as the surrounding temperature for the radiative transfer calculation. Agreement between thermocouple and IR temperatures within $\pm$ 0.05 °C is confirmed prior to testing. The mass of each water-filled test beaker (plus evaporator, when applicable) was weighed before and after each test period using the RADWAG PS 1000 electronic scale. Masses were not continuously monitored by the scale to avoid excess heat input from the scale, and to allow the two beakers to remain sufficiently close to one another to achieve identical ambient conditions. As the evaporation surface temperature drops very rapidly once the beakers are prepared during setup, the initial temperature of the bulk water used to fill each beaker is measured via thermocouple prior to each test, within 5 min of the start time. This temperature is used as the initial domain temperature in the analysis. Humidity was monitored during each test using the WS-2000 weather station with the WH32B sensor. The humidity during the comparative test was 18.6\ $\pm$ 2.1\%.  Differential scanning calorimetry curves were performed on the TA Instruments Q2000 DSC under Argon flow (50 ml min$^{-1}$) using a ramp rate of 5 °C min$^{-1}$. Thermal conductivity is characterized using the HotDisk TPS 2500s. SEM images were obtained from the Supra 25 SEM using an acceleration voltage of 5 kV.

\subsubsection*{Author contributions}
\textbf{Andrew Caratenuto:} Conceptualization, Methodology, Validation, Investigation, Writing -- Original Draft, Visualization; \textbf{Yi Zheng:} Conceptualization, Resources, Writing -- Review \& Editing, Supervision, Funding acquisition.

\subsubsection*{Acknowledgements}
The authors would like to thank Dr. Randall Erb and Daniel Braconnier for facilitating the differential scanning calorimetry measurements. This project is supported by the National Science Foundation through grant number CBET-1941743.

\subsubsection*{Conflicts of interest}
The authors declare no competing interests.

\bibliography{references}

\end{document}


\title[Critical assessment of water enthalpy characterization through dark environment evaporation]{\begin{center}
    \textbf{Supplementary information:}
\end{center} Critical assessment of water enthalpy characterization through dark environment evaporation}


\author[1]{\fnm{Andrew} \sur{Caratenuto}}

\author[1,2]{\fnm{Yi} \sur{Zheng}}\email{y.zheng@northeastern.edu}

\affil[1]{\orgdiv{Department of Mechanical and Industrial Engineering}, \orgname{Northeastern University}, \orgaddress{\city{Boston}, \state{Massachusetts}, \country{USA}}}

\affil[2]{\orgdiv{Department of Chemical Engineering}, \orgname{Northeastern University}, \orgaddress{\city{Boston}, \state{Massachusetts}, \country{USA}}}


\maketitle
\begin{figure} 
    \centering
    \includegraphics[width=\textwidth]{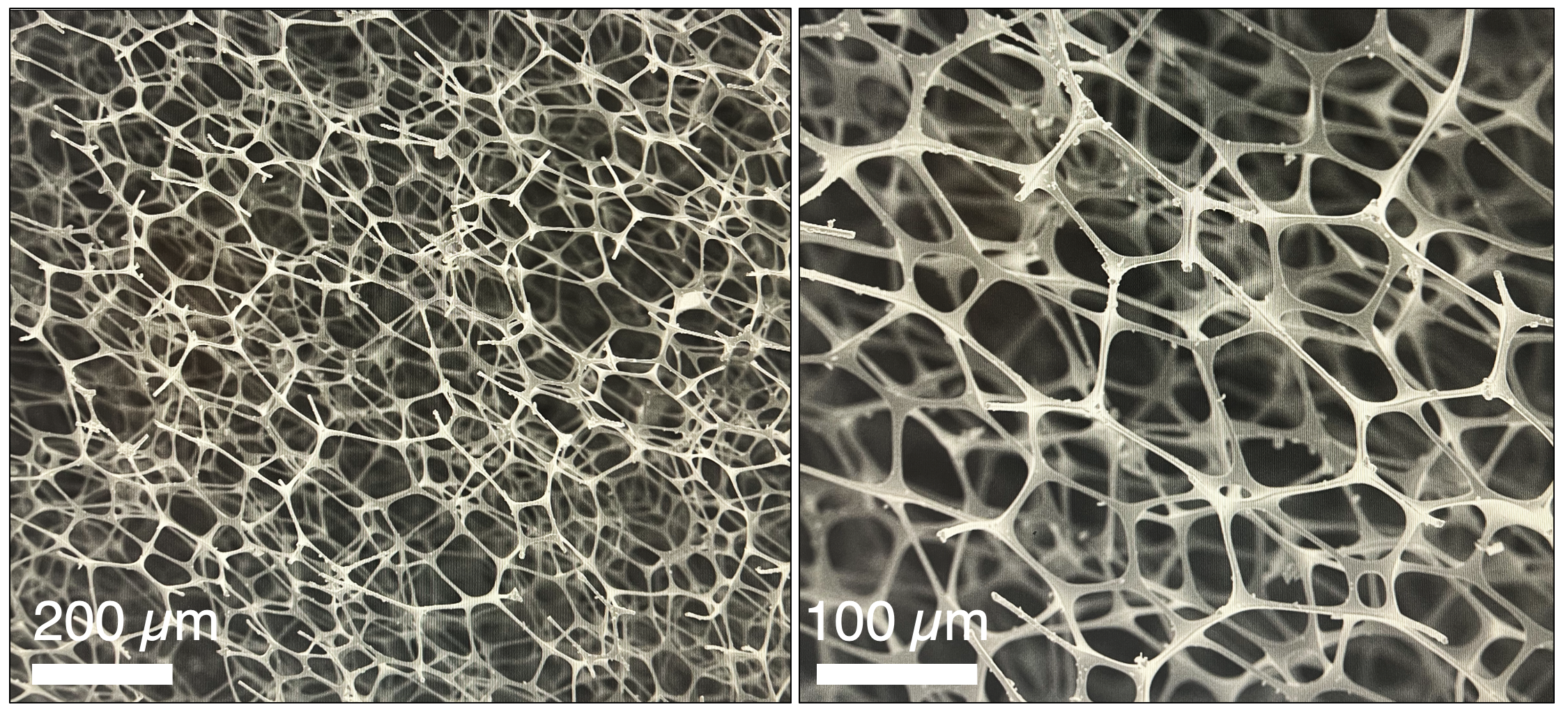}
    \caption{{SEM images of MF sponge evaporator.}}
    \label{f1}
\end{figure}

\begin{figure} 
    \centering
    \includegraphics[width=\textwidth]{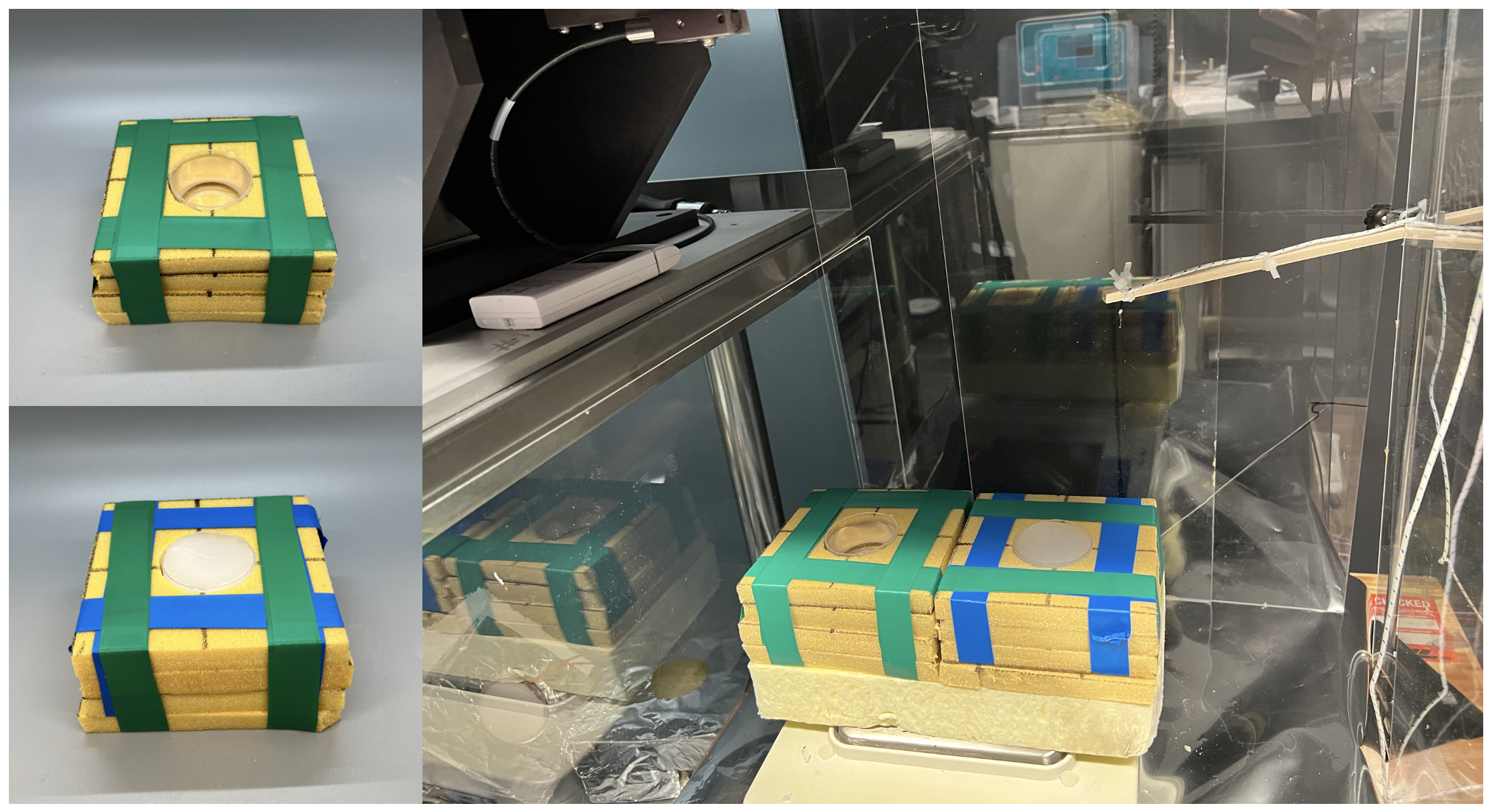}
    \caption{{Comparative dark environment experimental test setup.}}
    \label{setup}
\end{figure}

\section*{Supplementary Note 1}
\textbf{Transient heat transfer model}\\
Using the entire beaker as a control volume, the transient energy balance for each beaker is

\begin{equation}
    q_{evap}=\dot{m}h_{fg}=q_{conv}+q_{rad}+q_{cond}+q_{sens}.
    \label{ebalance}
\end{equation}
In this equation, $q_{evap}$, $\dot{m}$, $h_{fg}$, $q_{conv}$ and $q_{rad}$ maintain their definitions from equations in the main text. Conduction and sensible heat contributions are not neglected in this analysis for accuracy. Conductive heat transfer manifests from side and bottom heat fluxes through the insulation layers. The geometry of the beaker within the insulation block is a cylinder within a square; thus, typical plane wall or radial conduction treatment is not appropriate. Instead, the conduction shape factor $S$ is used \cite{incropera_fundamentals_2007}:

\begin{equation}
    S = \frac{2 \pi H}{\ln(\frac{1.08w}{d})},
    \label{S}
\end{equation}
where $w$ is the width of the insulating block, and $H$ and $d$ are the height and diameter of the beaker, respectively. With the relation $q_{cond} = Sk_{ins}\Delta T$, where $k_{ins}$ is the thermal conductivity of the insulating block, the heat rate due to conduction can be determined. However, the insulation blocks are placed side-by-side, such that one side of the insulation blocks are touching (Fig. S\ref{setup}). We assume that no conductive heat transfer occurs from this face due to the lack of contact with the ambient air. For this reason, we approximate the entire conduction contribution as $q_{cond} = 0.75Sk_{ins}\Delta T$ to account for this condition. In reality, this condition induces a small asymmetric contribution in the experimental case. However, the magnitudes of the heat transfer rates are sufficiently small, such that this asymmetry is not expected to make a significant contribution. This is also evidenced by IR camera images, which show a symmetric temperature distribution on the evaporation surface. For these reasons, the conductive heat flux on the boundary of the cylinder is set as 

\begin{equation}
    q''_{cond} = 0.75Sk_{ins}\frac{T_{\infty}-T_i}{\pi d H},
    \label{cond1}
\end{equation}
where the outer temperature of the insulation is assumed to be equal to the ambient temperature. For the bottom surface of the beaker, the standard plane wall conduction equation is used due to the parallel geometry \cite{incropera_fundamentals_2007}: 

\begin{equation}
    q''_{cond} = k_{ins}\frac{T_{\infty}-T_i}{L \pi (d/2)^2},
    \label{cond2}
\end{equation}
where $L$ is the bottom insulation layer thickness.

The sensible heat contribution $q_{sens}$ is expressed as  \cite{incropera_fundamentals_2007}:

\begin{equation}
    q_{sens} = \rho V c_{p}\frac{\Delta T_{mean}}{\Delta t}.
    \label{sens}
\end{equation}
The water density, specific heat, and beaker volume are given by $\rho$, $c_{p}$, and $V$, respectively. $\Delta T_{mean}$ is the difference in the bulk mean temperature of water over the corresponding time step $\Delta t$. A time step of 1 min is used, which is verified as sufficiently small to achieve convergence. 

The boundary condition at the top surface for each time step is set as the experimentally measured surface temperature of the evaporation surface. For the rest of the domain, the initial temperature is set as the bulk water temperature of water used to fill the beakers, measured just before the start of the experiment. The side and bottom walls are designated with conductive heat flux boundary conditions based on Eqs. \ref{cond1} and \ref{cond2}, respectively. Insulation thermal conductivity and thickness are set based on the experimentally measured values of the surrounding insulation, as given in \textit{Experimental Methods}, and temporal data is used for the ambient temperature. Water properties $k$, $c_p$, and $\rho$ are set as 0.6 W m$^{-1}$ K$^{-1}$, 4182 J kg$^{-1}$ K$^{-1}$, and 999 kg m$^{-3}$, respectively \cite{moran_fundamentals_nodate}. Changes in volume and/or mass are not considered based on the small evaporated mass over the course of the experiment.

\clearpage

\begin{figure} 
    \centering
    \includegraphics[width=\textwidth]{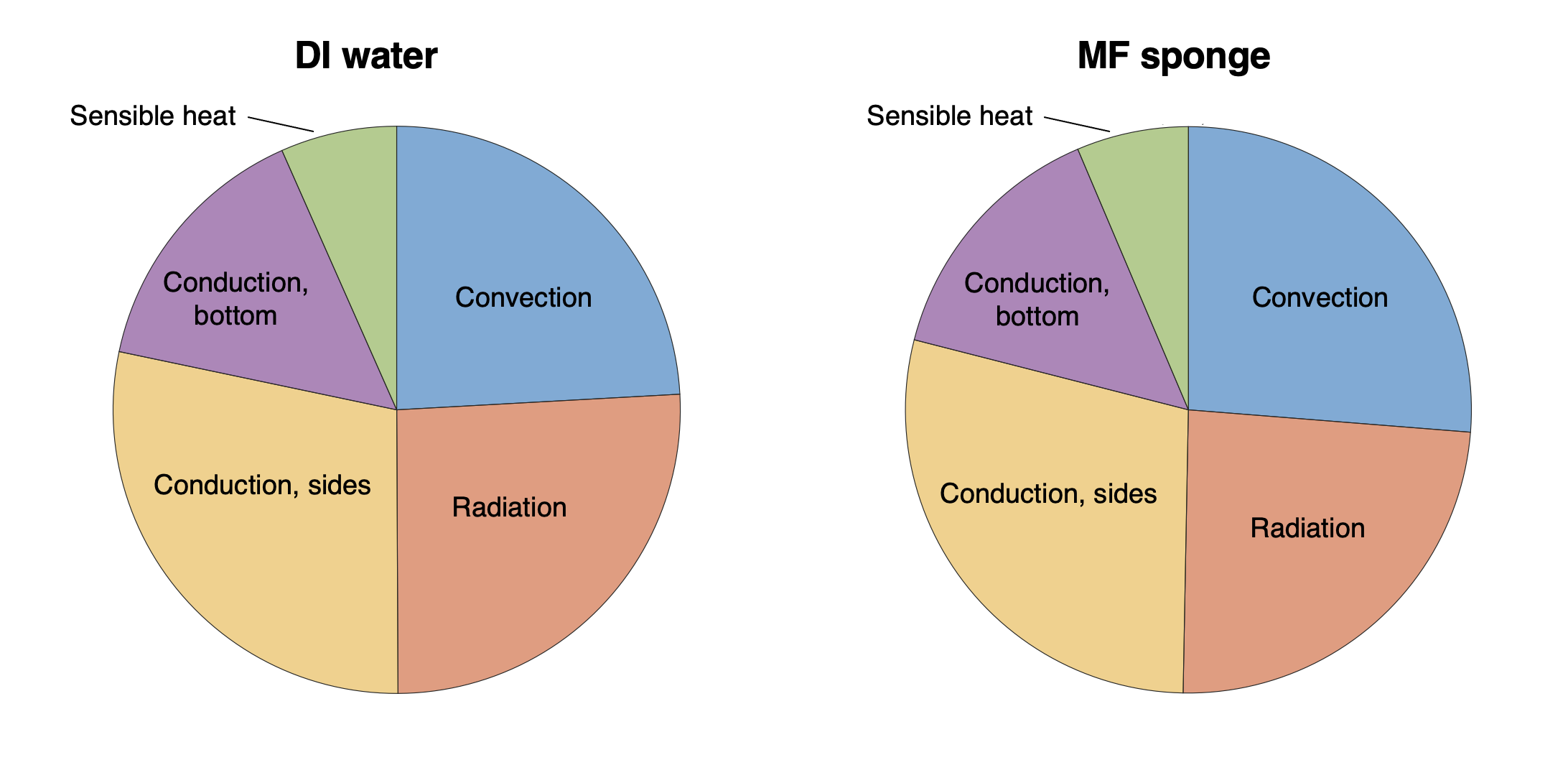}
    \caption{{Distribution of heat transfer mechanisms for each evaporation system.}}
    \label{f2}
\end{figure}

\begin{figure} 
    \centering
    \includegraphics[width=\textwidth]{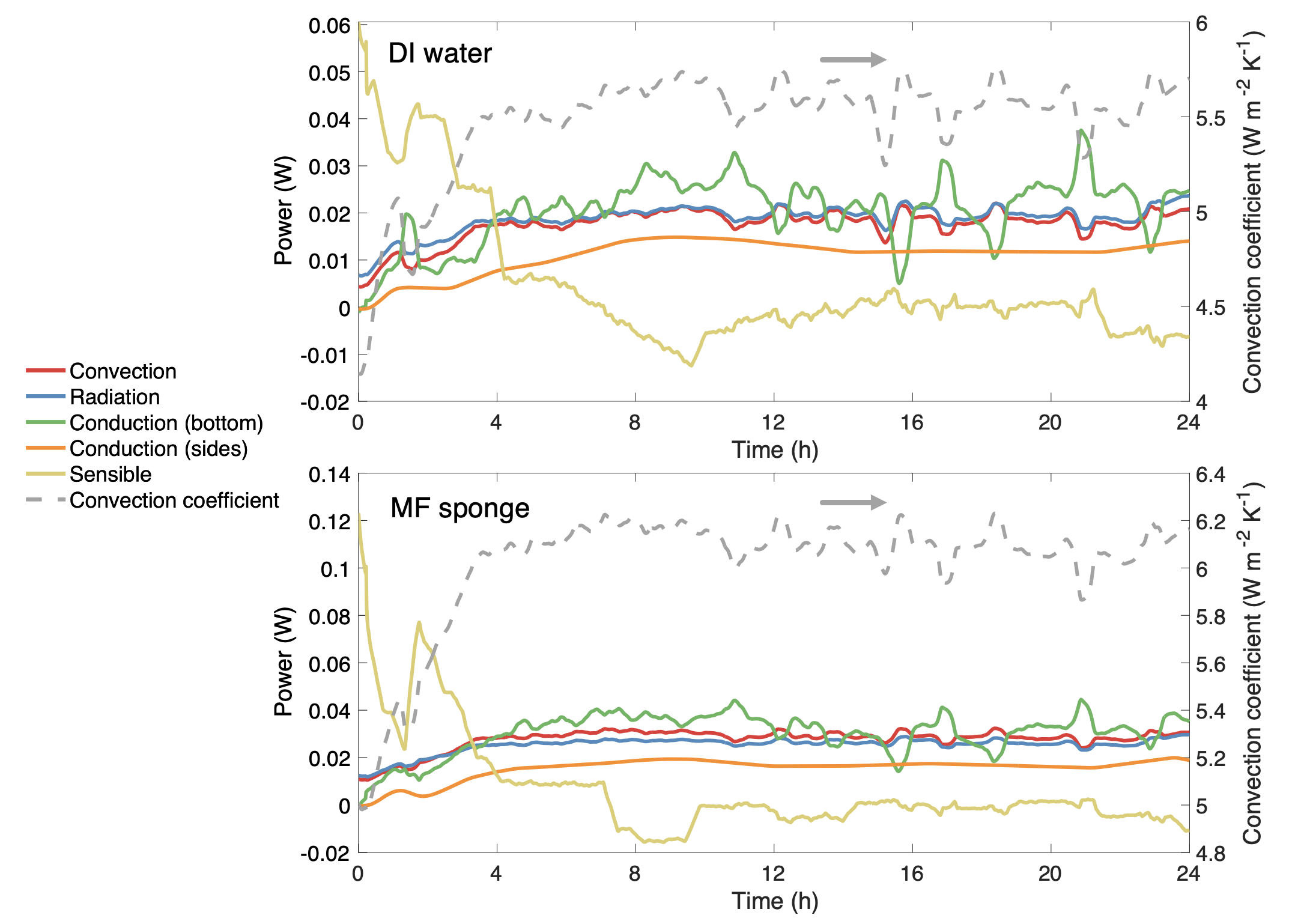}
    \caption{{Time-dependent heat transfer mechanisms for each evaporation system during the 24-h test.}}
    \label{f2}
\end{figure}

\clearpage

\section*{Supplementary Note 2}
\begin{figure} 
    \centering
    \includegraphics[width=\textwidth]{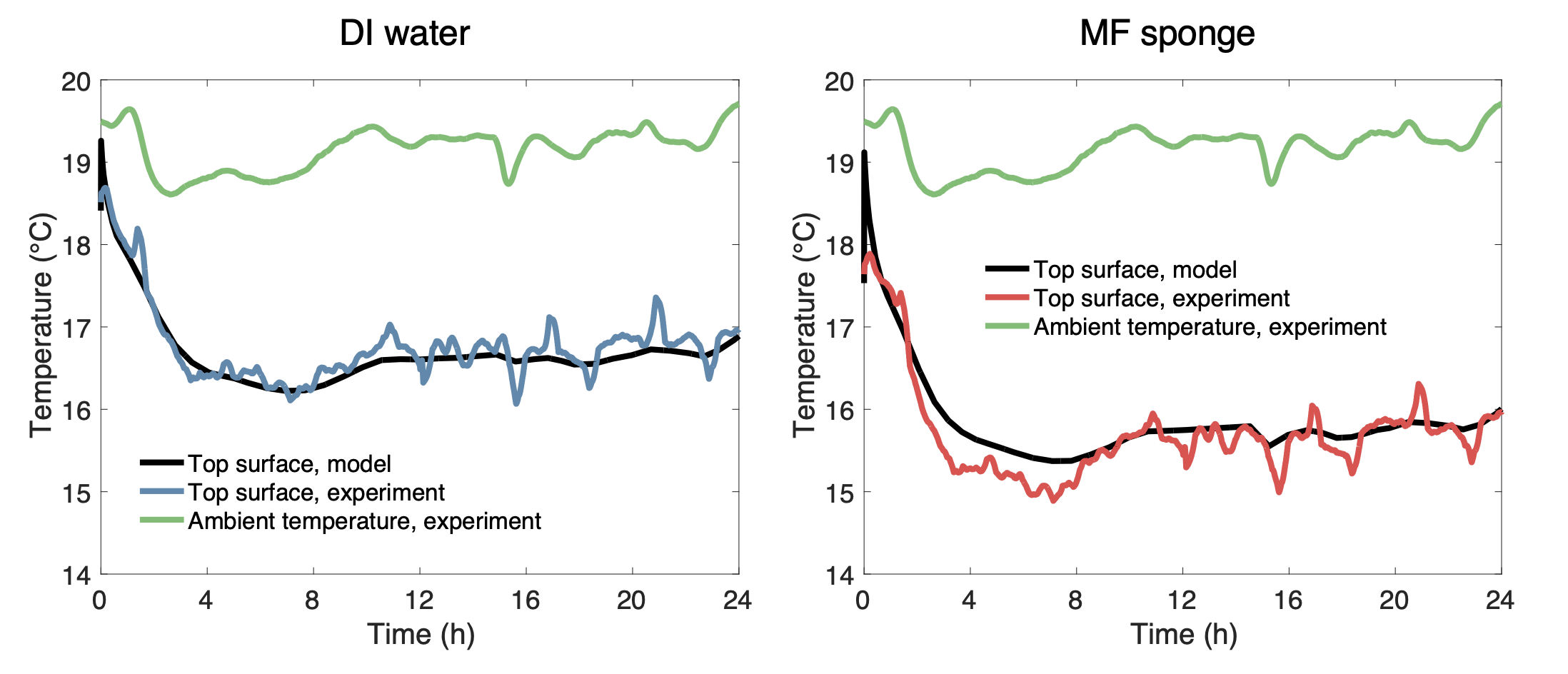}
    \caption{Alternative model results, using the evaporation boundary condition as described in Supplementary Note 2.}
    \label{evap2}
\end{figure}

\textbf{Evaporation boundary condition at top surface}\\
To further support our experimental and modeling results, an alternative version of the transient model is simulated, as shown in Fig. S\ref{evap2}. In this model, the side and bottom boundary conditions are identical to those described in the main text, as are the given water properties and the initial temperatures of the beaker. The top boundary condition is modified as a heat flux condition based on the evaporative heat flux. Because convection and radiation also occur at this surface, the top surface heat flux boundary condition is set as:

\begin{equation}
    q''_{top}=\frac{q_{conv}+q_{rad}-q_{evap}}{A_{top}}.
    \label{ebalance}
\end{equation}

Equations for convective, radiative, and evaporative heat fluxes are identical to those given in the main text. The evaporative heat flux of each beaker is assumed to be a constant. Its value is obtained based on the 24-h averaged evaporation rate and a vaporization enthalpy value of 2460 J g$^{-1}$. The result of this model provides the temperature distribution within the beaker, which allows for the comparison of the experimentally-measured top surface temperature with the result of the model. Both the DI water and MF sponge cases show relatively good agreement with the experimental result, maintaining similar responses to modifications in the ambient temperature. This provides an additional method to illustrate that the measured evaporation rates and temperature data are reliable, as they are physically consistent with the energy balance in multiple versions of this model. Deviations from the measured temperatures result from several factors. For one, the evaporation rate is not likely constant throughout the test, and may change slightly based on ambient humidity and temperature variations. In addition, evaporation will actually begin prior to this test period during setup, although this setup time is minimized (kept to 5 min or less in practice). Also, the top surface is initialized as the first experimental temperature measurement, while the rest of the beaker is initialized at the measured bulk water temperature. In combination, these latter two factors result in a sharp increase in surface temperature, followed by a sharp decrease thereafter. However, both of these effects dissipate in the very early stages of the simulation.

\clearpage

\bibliography{references}